\documentclass[conference,10pt]{IEEEtran}
\IEEEoverridecommandlockouts
\usepackage{graphicx} 
\usepackage{cite}
\usepackage{xcolor}
\usepackage{amssymb}
\usepackage{amsmath}
\usepackage{commath}
\usepackage{bbm}

\newcommand{\RFI}{\text{RFI}}
\newcommand{\RFImax}{\text{RFI}_{\text{max}}}

\title{Reducing Satellite Interference to Radio Telescopes Using Beacons}
\author{Cuneyd Ozturk,\thanks{C.~Ozturk, R.~A.~Berry, D.~Guo, and M.~L.~Honig are with the Department of Electrical and Computer Engineering, Northwestern University, Evanston, IL, 60208 USA (e- mail: \{cuneyd.ozturk, rberry, dguo, mhonig \}@northwestern.edu)} Randall A.~Berry, Dongning Guo, Michael L. Honig, Frank D. Lind\thanks{F.~Lind is with Haystack Observatory, Massachusetts Institute of Technology, Cambridge, MA, 02139, USA (e-mail: flind@mit.edu)} \thanks{This work was supported by the U.S. National Science Foundation (NSF) through SpectrumX, an NSF spectrum innovation center.} }
\begin{document}
\maketitle

\begin{abstract}
%\dg{SUGGEST SAYING SOMETHING LIKE ``we propose to use beacons'' TO INDICATE ORIGINALITY.}
This paper proposes the transmission of beacon signals to alert potential interferers of an ongoing or impending passive sensing measurement. We focus on the interference from Low-Earth Orbiting (LEO) satellites to a radio-telescope. We compare the beacon approach with two versions of Radio Quiet Zones (RQZs): fixed quiet zones on the ground and in the sky, and dynamic quiet zones that vary across satellites. The beacon-assisted approach can potentially exploit channel reciprocity, which accounts for short-term channel variations between the satellite and radio telescope.
System considerations associated with beacon design and potential schemes for beacon transmission are discussed. The probability of excessive Radio Frequency Interference (RFI) at the radio telescope (outage probability) and the fraction of active links in the satellite network are used as performance metrics. Numerical simulations compare the performance of the approaches considered, and show
that the beacon approach enables more active satellite links relative to quiet zones for a given outage probability. 
\end{abstract}
\section{Introduction}
Radio telescopes used for astronomy (i.e. passive sensing) are typically very sensitive, as they must detect extremely weak signals from distant, faint astronomical sources\cite{2021_ITU_R_2259, 2014_Carol_PropPrediction,2021_Minn_NGSO_Ras, 2023_Chakraborty_CellularRFI, 2022_Weldegebriel_Pseudonymetry}. The larger telescopes may utilize highly specialized receivers with low-noise amplifiers to amplify weak signals and cryogenic cooling to reduce thermal noise to achieve ultra-high sensitivity\cite{2013_Bryerton_Cyrogenic}. A modern radio telescope can be 
150 dB more sensitive than a GSM cell phone\cite{2014_Carol_PropPrediction, 2021_ITU_R_2259}. 

The high sensitivity of radio telescope receivers makes them susceptible to radio frequency interference (RFI) from other sources, such as radio and television transmitters, mobile phones,  satellites,  and other wireless devices\cite{2014_Carol_PropPrediction, 2021_ITU_R_2259}.
RFI can limit the precision of radio telescope measurements and obscure or distort the weak signals from astronomical sources \cite{2021_ITU_R_2259, 2022_ITU_R_2188}.  Particularly, when RFI signals are strong enough to drive the amplifiers in a radio telescope receiver system into saturation or the non-linear regime, radio observations cannot be calibrated precisely and no useful data is obtained. For useful radio frequency observations, any RFI must permit linear operation of the amplifiers \cite{2021_ITU_R_2259, 2022_ITU_R_2188}.

The International Telecommunications Union (ITU) has implemented regulations pertaining to the allocation and utilization of spectrum in order to effectively manage and mitigate potential conflicts arising from various services \cite{2023_Zheleva_RDZ, 2023_Minn_SRQZ}. Radio quiet zones (RQZs) have been designated to protect scientific users from interference due to active users that may operate in adjacent bands \cite{2021_ITU_R_2259, 2023_Zheleva_RDZ}.  Defining the boundaries of RQZs presents a problem involving a trade-off between passive and active users. RQZs must be sufficiently large so that RFI will not disrupt measurements made by the radio telescope. On the other hand, the area should not be too large so as to unduly limit spectrum utilization by active users \cite{2014_Altamimi_Enforcement}. 
 
 In 1958, the United States established the first RQZ at the Green Bank Observatory station for the protection of radio astronomy observations \cite{2021_ITU_R_2259}.  In accordance with that original RQZ, multiple RQZs have been established in various locations globally. Examples include the RQZs surrounding the Large Millimeter Telescope in Mexico, the ALMA telescopes in Chile, and the RQZ in Western Australia \cite{2021_ITU_R_2259}. 

 The majority of early transmitters, with the exception of amateur radio, were operated and controlled by government agencies or large corporations. In contrast, in recent years the vast majority of low-powered transmitters have operated within the private sector \cite{2021_ITU_R_2259}. This complicates the evaluation and management of radio transmitters near an RQZ. In addition, early devices were scarce and, in the majority of cases, fixed. Recent tendencies have shifted towards much higher densities of mobile devices and networks that can be rapidly deployed. Moreover, the international nature of low earth orbit (LEO) satellite mega-constellations will compromise the effectiveness of the RQZ protections \cite{2021_ITU_R_2259}.

Concern about spectrum conflicts between scientific and active users has increased significantly, despite the fact that these technologies differ greatly in terms of sensitivity levels, interference tolerance, and space, time, and frequency usage patterns \cite{2021_ITU_R_2259, 2023_Zheleva_RDZ, 2023_Chakraborty_CellularRFI, 2022_Weldegebriel_Pseudonymetry}.  Widespread use of different broadband technologies, including LEO satellite mega-constellations, are likely to increase and intensify conflicts \cite{2021_ITU_R_2259, 2016_Minn_SpecSharingCellular, 2021_Minn_NGSO_Ras, 2022_Weldegebriel_Pseudonymetry}. Since active users cannot determine exactly when scientific users are taking measurements via spectrum sensing, dynamic spectrum access techniques for coordinating active users have not been 
directly applied to resolve conflicts \cite{2022_Weldegebriel_Pseudonymetry}.

\subsection{Main Results}
We discuss potential methods for enabling the LEO satellite networks to share spectrum resources over time and space with radio astronomy. We start by presenting two different versions of the RQZ. We first consider quiet zones both in the sky and the ground with fixed radii (Fixed Quiet Zones). We subsequently propose quiet zones on the ground which vary based on the satellite location (Dynamic Quiet Zones). We present a simple path-loss model that provides insight into the tradeoff between the probability of excess RFI at the radio telescope versus number of active satellite links. The model accounts for beam patterns at both the radio telescope and the satellites.

Our analysis shows that the fixed quiet zones approach may prove overly restrictive for the satellite network. By incorporating antenna orientations into the design through the utilization of dynamic quiet zones, a significant improvement in the number of active satellite links is attained.
%\mh{This is not a conclusion from the analysis but rather is an assumption for the methods. What did you learn from the analysis?}

As an alternative to varying RQZs dynamically, we propose an approach where the radio telescope or nearby antenna transmits a beacon signal that alert satellites when and where passive measurements are being made (i.e. frequencies, look directions, integration times, etc). A motivation is to exploit near reciprocity so that successful decoding of the beacon at the satellite according to a preset threshold implies that it would generate excessive RFI at the radio telescope. In that scenario the satellite must cease any transmissions. Adjusting the beacon power then controls the extent of the RQZ and number of active satellite links. We extend the preceding path-loss model to analyze the probability of excessive RFI as a function of beacon power, number of satellite links, and beam characteristics.  
Our analysis shows that because the beacon approach can account for short-term channel variations, more satellite links can be enabled compared with the RQZ approaches while achieving the same outage probability.

%\dg{THIS WAY OF INTRODUCING BEACONS FEELS VERY CASUAL. IT IS THE CORE IDEA HERE. SHOULD MAKE THE ORIGINALITY OBVIOUS TO THE READER. ALSO, IT IS WORTH PROVIDING A SUMMARY OF OUR FINDINGS IN THE INTRODUCTION. THE READER SHOULD BE ABLE TO FIND OUT WHAT'S GOOD/BAD ABOUT WHAT WE PROPOSE. SOME READERS WILL READ NO FURTHER THAN THE INTRO.}

\section{System Model}

\begin{figure}
    \centering
    \includegraphics[width = \linewidth]{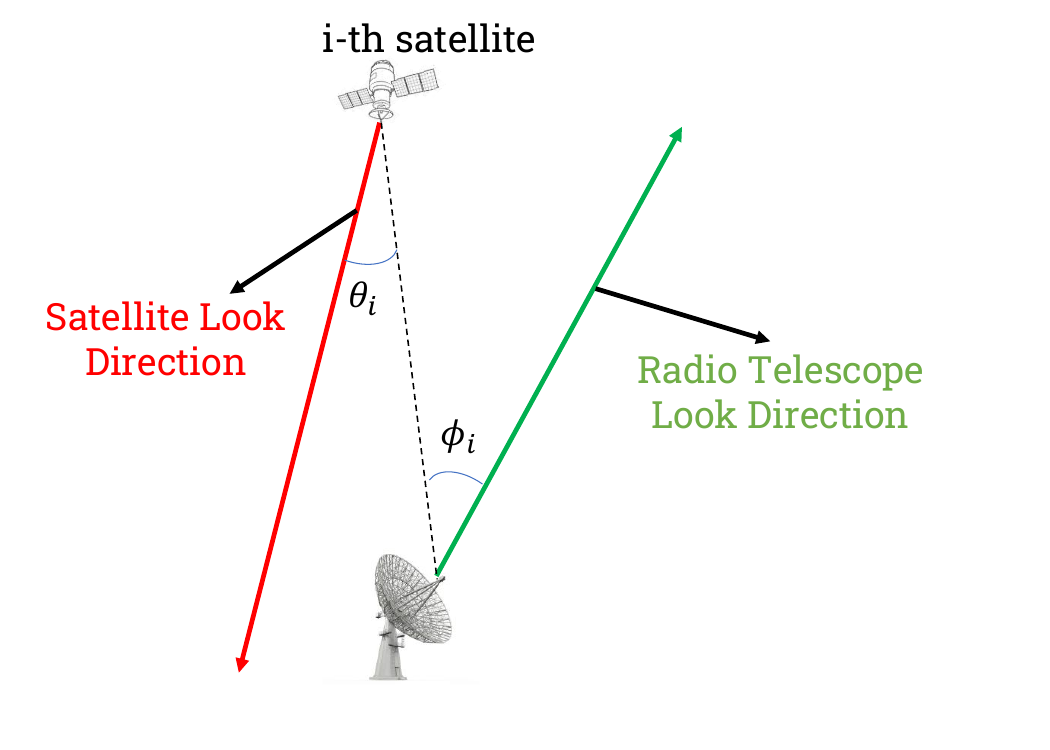}
    \caption{Off-axis angles for the $i$-th satellite and the radio telescope.}
    \label{fig:sat_geo}
\end{figure}
A passive sensor such as a radio telescope in general conducts a measurement by collecting received energy over a particular frequency band $B$ for a duration of $T$ seconds along a designated look direction. A transmitter in or adjacent to the passive sensing band generates RFI at the radio telescope.  In this paper we consider RFI from satellites, where the arrival of the satellites to the field-of-view of the radio telescope is modeled as a stochastic process. The aggregate RFI level at the radio telescope is denoted as $\RFI(t)$ for $t \in [0, T]$.  The maximum allowable level of RFI at the radio telescope is denoted as $\RFImax$.  The radio telescope discards its recorded data that corresponds to time instances when the instantaneous RFI level exceeds $\RFImax$.

The satellites are LEO, and are positioned at an altitude of $h = 550~$km which corresponds to Starlink LEO satellite constellation \cite{cakaj2021parameters}. Moreover, our methods are not limited to this specific altitude. In particular, satellites reside on the surface of the sphere centered at $(0, 0, 0)$ with the radius $R_e + h$, where $R_e = 6371$~km is the radius of the earth. Let $N$ represent the number of distinct satellites that enter the field-of-view of the radio telescope throughout the entire measurement interval $[0, T]$.  If the value of $T$ is sufficiently large, $N$ can be as many as the total number of satellites within the constellation, although at a particular time, only a fraction of those will be in the field-of-view. 

We consider the worst case scenario where the satellites and the radio telescope are using the same frequency band. For adjacent channel interference, our method still applies, but the path loss between satellites and the radio telescope must be adjusted to reflect the attenuation by filtering of the adjacent RFI. 

We denote $p_i(t)$ as the transmit power of the $i$-th satellite at time $t$; we denote $d_i(t)$ as the distance between the $i$-th satellite and the radio telescope at time $t$. The Free Space Loss (FSL)  between the $i$-th satellite and the radio telescope is expressed as
~\cite{2021_ITU_R_2259, 2019_ITU-P_525}:
\begin{align}\label{eq:FSL}
    \text{FSL}\left(d_i(t)\right) = \left(\frac{4\pi d_i(t) f_c}{c}\right)^2,    
\end{align}
where $c$ is the speed of the light and $f_c$ is the carrier frequency.

As depicted in Fig.~\ref{fig:sat_geo}, we use $\theta_i(t)$ to denote the angle between the look direction of the $i$-th satellite and the direction from the $i$th satellite to the the radio telescope at time $t$;  we use $\phi_i(t)$ to denote the  angle between the look direction of the radio telescope and the direction from the radio telescope and the $i$th satellite at time $t$. We denote $w_T(\cdot)$  as the  antenna pattern of the satellites (assuming all satellites have the antenna pattern), and $w_R(\cdot)$ as the antenna pattern of the radio telescope.

The fundamental problem is enabling more efficient use of spectrum across time and space accounting for both the radio telescope and the satellite network. In other words, we would like to determine a set of satellite links to activate, subject to constraining the RFI level to be within the radio telescope's tolerance, which may depend on the particular measurement. 

Let $H_t$ represent the set of active satellites that are positioned over the field-of-view of the radio telescope for any given time $t\in[0,T]$. Then, for $i\in H_t$, the interference due to the $i$-th satellite at time $t$ is modeled as 
\begin{align}
    I_i(t)  = p_i(t) \frac{w_T\left(\theta_i(t)\right) w_R\left(\phi_i(t)\right)}{\text{FSL}\left(d_i(t)\right)} \xi_i(t),
\end{align}
where $\xi_i(t)$ captures the uncertainty in the instantaneous RFI level. We model the $\xi_i(t)$'s as independent stochastic processes across the satellites. The aggregated $\RFI$ at time $t$ can be expressed as
%\dg{SUGGEST REPLACING Ht BY A SET WE CONTROL TO OBTAIN A GENERAL EXPRESSION}
\begin{align}
    \RFI(t) = \sum_{i\in H_t} I_i(t).
\end{align}
We define the outage probability as the average fraction of the time that the aggregated RFI level exceeds $\RFImax$,
\begin{align}\label{eq:outage}
    P_{\text{out}}\triangleq \mathbb{E}\Bigg[\frac{1}{T}\int_{0}^{T} \mathbbm{1}\{\RFI(t)\ge \RFImax\} \, dt\Bigg],
\end{align}
where $\mathbbm{1}\{C\}$ denotes the indicator of condition $C$, i.e., $\mathbbm{1}\{C\}=1$ if condition $C$ holds; otherwise, $\mathbbm{1}\{C\}=0$. In general, one should note that the powers and the locations of the satellites may be time-varying.

For simplicity, we will use the number of active satellite links as the performance metric for the satellite network. In the subsequent sections, we discuss two distinct methods for allocating the spectrum across the satellites over time.
\section{Radio Quiet Zones}\label{sec:RQZ}
\begin{figure}
    \centering
    \includegraphics[width = 1\linewidth]{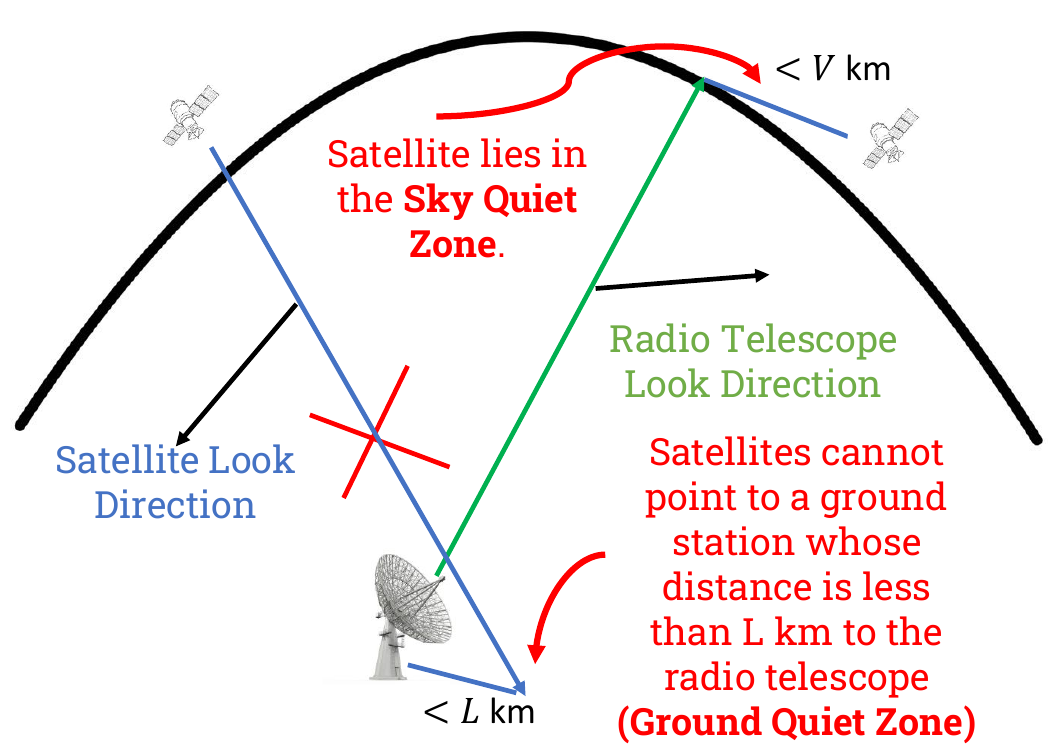}
    \caption{Illustration of the fixed quiet zone (sky quiet zone and ground quiet zone).}
    \label{fig:quietzones}
\end{figure}
In this section we present two different models for designing RQZs.
\subsection{Fixed Quiet Zones}
We consider quiet zones  both on the ground and in the sky. In particular, the space is partitioned into regions as illustrated  in Fig.~\ref{fig:quietzones}:
\begin{itemize}
    \item Ground Quiet Zone: This is a region around the radio telescope. If the distance between a ground station and the radio telescope is less than $L$~km, the ground station will not receive any satellite service.
    \item Sky Quiet Zone: This region is based on the intersection point of the look direction of the radio telescope and the sphere on which the satellites reside. If the distance of a satellite to the intersection point is less than $V$~km, the satellite cannot transmit.    
\end{itemize}
The radii $L$ and $V$ are taken as design parameters. Both the outage probability in \eqref{eq:outage} and the number of active satellite links are functions of $V$ and $L$. The optimal values of $V$ and $L$ achieve a desired outage probability while minimizing the number of deactivated links. 

The simplicity of this approach is its key benefit. $V$ and $L$ can be computed offline and announced to the satellites. For instance, this announcement could be transmitted through a terrestrial connection to a satellite ground station.  The feedback from the radio telescope does not vary over short time scales. However, this design is rigid in that it disregards the satellite antenna orientations: a satellite located in the sky quiet zone may not create harmful interference at the radio telescope if its look direction is pointed away from the satellite. In addition, the footprints of the satellite transmitters may span hundreds or thousands of kilometers, making it difficult to restrict the quiet zones without affecting much larger areas \cite{2021_ITU_R_2259}.

\subsection{Dynamic Quiet Zones}
\begin{figure}
    \centering
    \includegraphics[width = 1\linewidth]{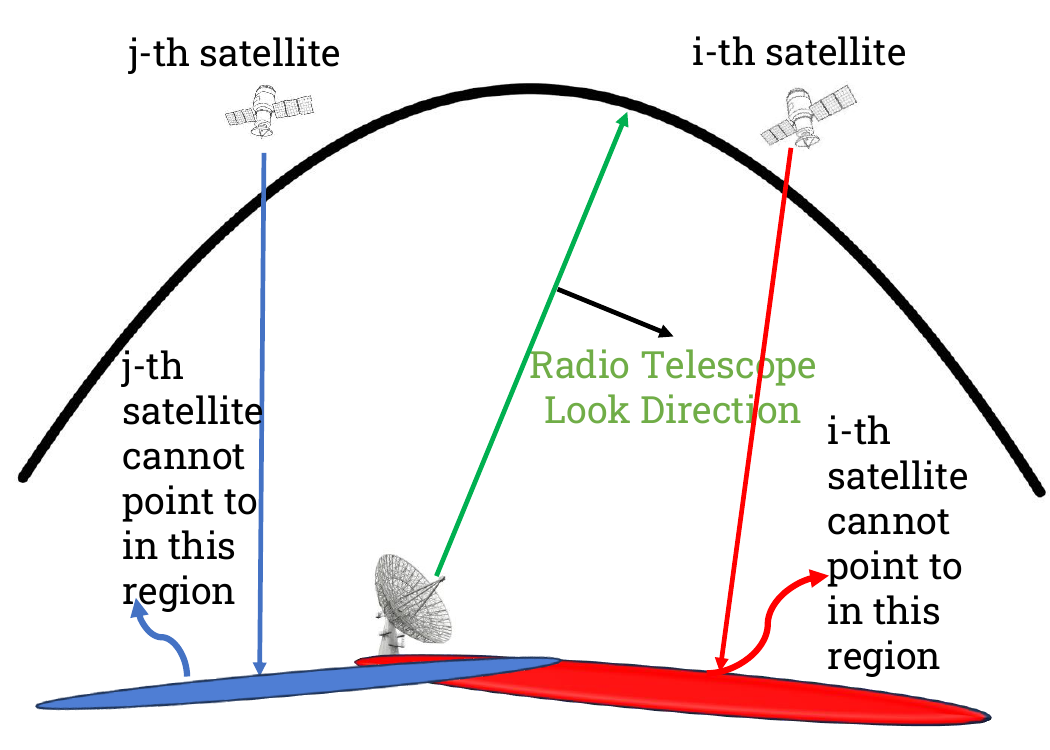}
    \caption{Dynamic quiet zones for two different satellites.}
    \label{fig:varying_quietzones}
\end{figure}
To increase the number of active satellite links with quiet zones, we consider varying the quiet zones dynamically.
%to account for the look directions of the satellites.
Referring to Fig. \ref{fig:varying_quietzones}, each satellite can autonomously determine the region on the ground to which they cannot point based on their transmit power profiles and frequency assignments given an RFI threshold $\tau$. These regions then change across satellites, depending on their locations. 

One drawback of this method is that it requires significantly more computation than the fixed quiet zones method. If the number of satellites seen by the radio telescope and their trajectories are predictable over a particular time period (e.g., a few days), then the determination of quiet zone boundaries for all satellites might be performed periodically and reloaded into the memory of each individual satellite. Alternatively, a band manager or the radio telescope operator may perform these calculations and periodically communicate the results to satellites through an independent channel.

\section{Beacon-Assisted Protocols}\label{sec:Beacon}
\begin{figure}
    \centering
    \includegraphics[width = 0.9\linewidth]{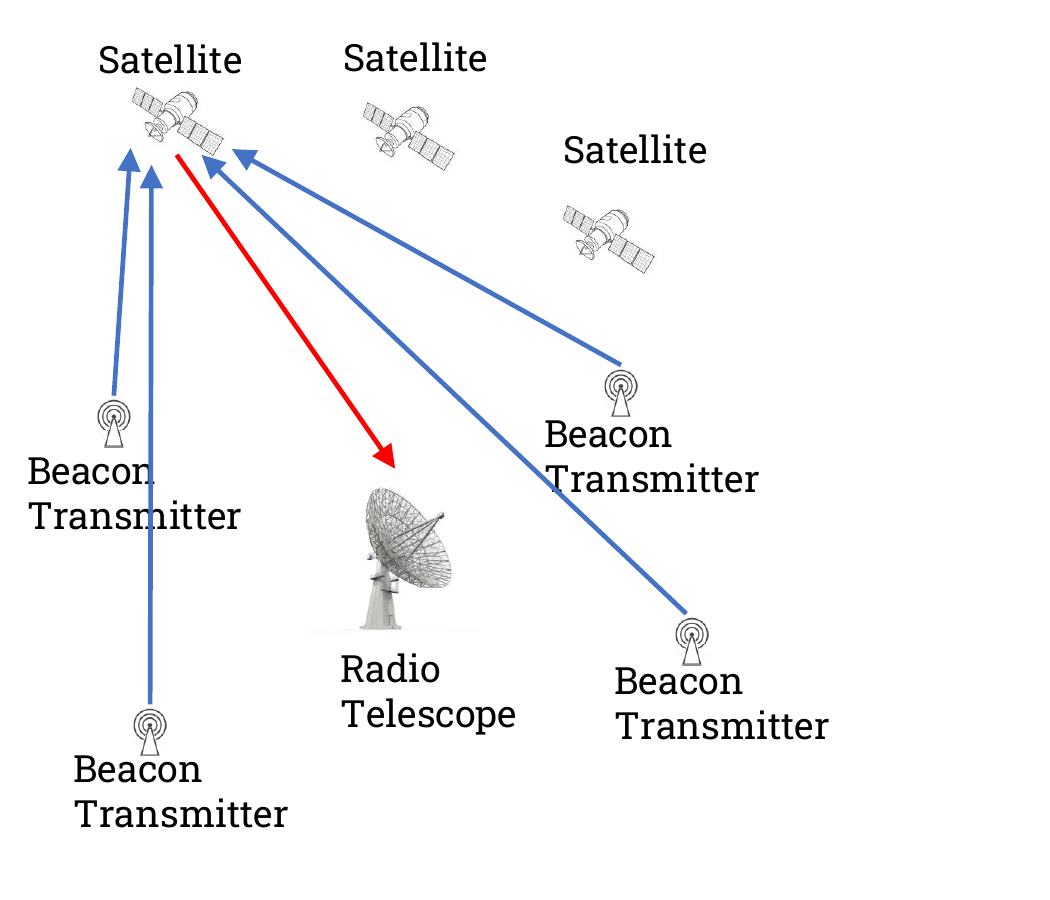}
    \caption{Illustration of the beacon approach.}
    \label{fig:beacon}
\end{figure}

Neither the fixed nor dynamic quiet zones methods capture short-term channel variations that may occur between the radio telescope and the satellites. As pointed out in \cite{2021_ITU_R_2259}, RQZs may include options for notification of active users. To that aim, we propose a method for establishing a dynamic communication environment between active and scientific users that enable more efficient use of spectrum resources. 

%\dg{In this section, we evaluate the proposal of letting the radio telescope operator send}
Referring to Fig. \ref{fig:beacon}, we consider the scenario in which the radio telescope can control the transmission of beacon signals to notify potential interferers about measurements in progress. The beacon transmitter(s) might be located at or near the radio telescope, and operate within or adjacent to the band being sensed. One option would be to exploit frequencies associated with satellite command and control uplink. Another might be to act as a user ground terminal which provides information via a data channel or via an added modulation or side channel. 
 
This approach can exploit instantaneous channel reciprocity if the beacon transmitter is located at the radio telescope. If the beacon transmitters are located near the radio telescope, the compromise in the channel state information may not be significant. That is, if a satellite can (cannot) detect the beacon, then it will (will not) likely interfere with the measurement. Hence beacons can take into account dynamically varying channel state information as well as variations in passive antenna operations and orientations. Moreover, the satellites need to equipped with the additional receivers to sense the beacon.

\subsection{Design Considerations}
 The beacon power profile may be spread over time, frequency and space. We denote $P_b(t, f, \Omega)$ as the beacon power profile for a given time $t$, frequency $f$, and spatial direction $\Omega$. The question we consider is how to design the beacon power profile $P_b(t, f, \Omega)$ to maximize the number of active links in %the spectral efficiency of 
 the satellite network subject to a constraint on the probability of excessive RFI at the radio telescope. %To simplify the problem we use number of active satellite links as a surrogate for spectral efficiency. 
 In this section we outline some basic design considerations.
 We assume that the satellites know when and where in frequency and space to listen for the beacon.

 \paragraph{Beacon duty cycle}  
 During the measurement, many satellites may enter and exit the radio telescope's field-of-view. Hence the beacon must be transmitted frequently enough to capture time-variations in the satellite locations and associated channels. One possibility is that the beacons are transmitted periodically and upon entering the field-of-view of the radio telescope, a satellite must wait for that period to ensure that there is no beacon before transmitting. Clearly, the cost of this approach is that it would impact the throughput of the satellites. On the other hand, the advantage of non-continuous transmission of beacons is reducing the possible RFI from the beacons to the radio telescope. For the subsequent analysis we assume that the beacon signal is continuously transmitted. That is, there is a dedicated frequency band $B_b$ for the beacon signal where the beacon signal is always present during the measurement.

\paragraph{Ultra-wideband signaling} This refers to an impulsive type of signal that periodically illuminates the entire channel. The primary disadvantage of this scheme is that the radio telescope may need to be turned off during the beacon transmission periods due to the large peak-to-average power of the beacon, which can drive the radio telescope into saturation or the non-linear regime. This is difficult in practice, due to recovery and hysteresis effects associated with dynamically shunting the beacon signal away from the radio telescope front-end.

\begin{figure}
    \centering
    \includegraphics[width = 1\linewidth]{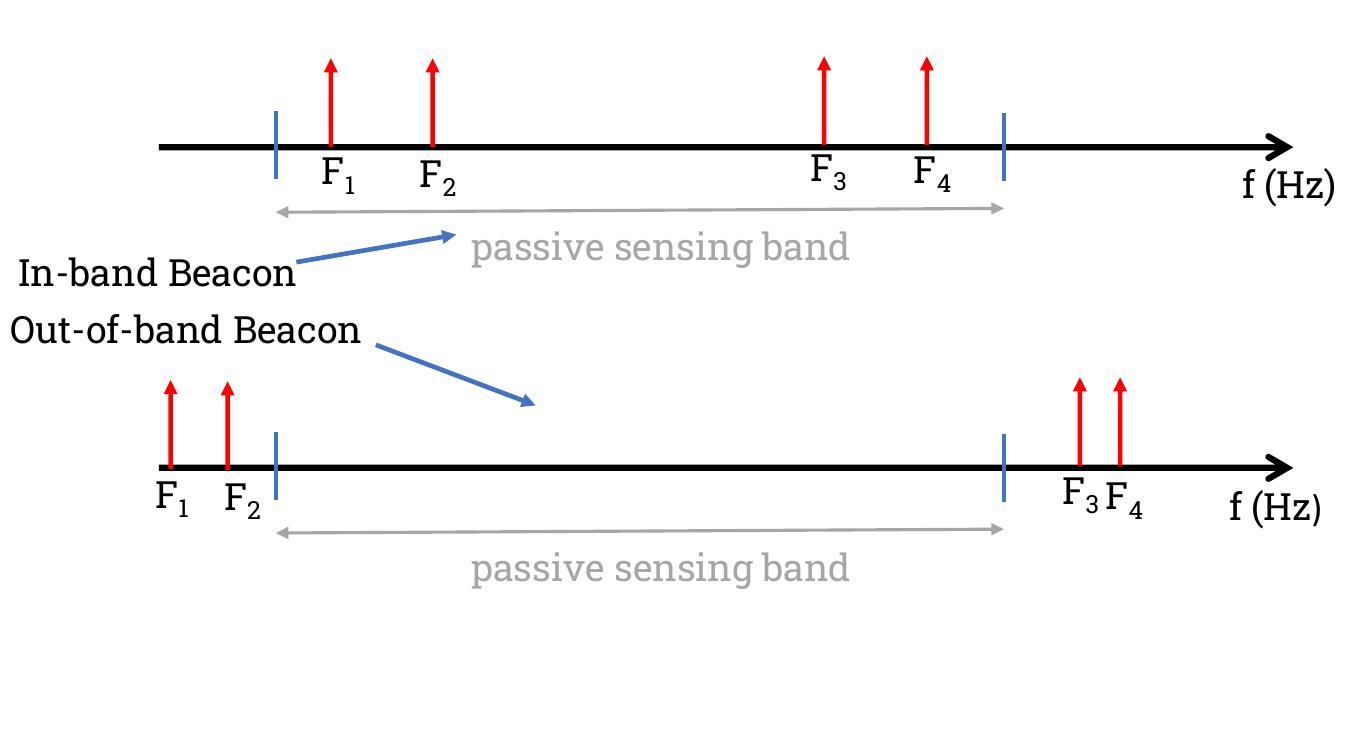}
    \caption{In-band and out-of-band beacons with four different carriers.}
    \label{fig:in-out-band}
\end{figure}

\paragraph{In-band versus out-of-band signaling}
Referring to Fig. \ref{fig:in-out-band}, the beacon signal might be transmitted within the passive sensing band (in-band), or in frequencies adjacent to the passive sensing band  (out-of-band).  
By exploiting channel reciprocity, an in-band beacon can account for frequency selectivity, but deletes the part of the spectrum that can be sensed, because of its presence. In contrast, an out-of-band beacon is subject to band-edge propagation, which may have different characteristics from in-band propagation, increasing the probability of excessive RFI. Exploitation of satellite control uplink channels would depend on the placement of the uplink relative to the passive observation band in question. 
%\dg{REMOVE SOURROUNDING BOX OF FIG. 4?}

\begin{figure}
    \centering
    \includegraphics[width = 1.1\linewidth]{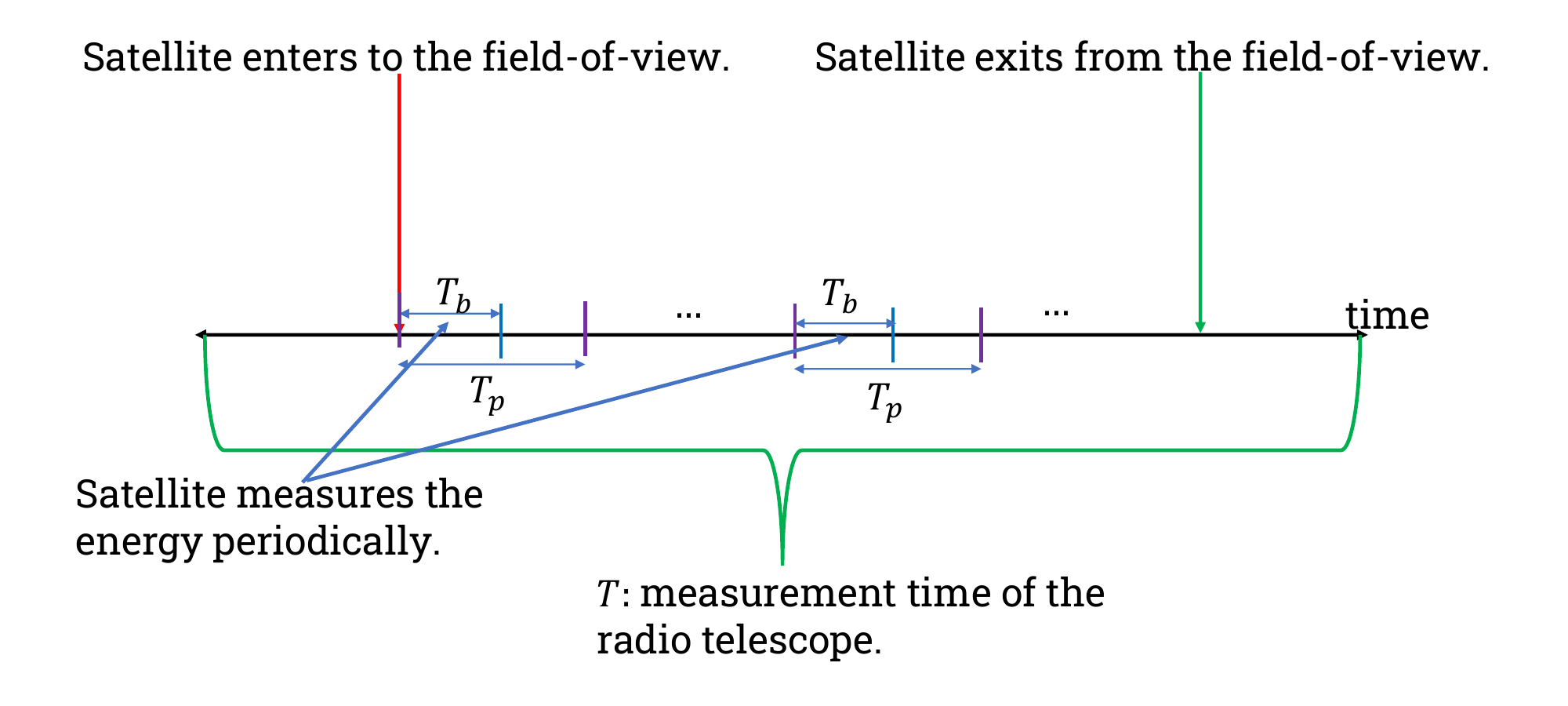}
    \caption{Beacon transmission scheme when the satellites periodically measure the energy of the beacon. %\dg{FONT IN GRAPH IS A BIT TOO SMALL.}
    }
    \label{fig:beacon_sensing}
\end{figure}

\paragraph{Spatial design} Another degree of freedom is how to spread the beacon power across space, i.e., the beacon antenna pattern. To exploit reciprocity, the beacon antenna pattern should be matched to that of the radio telescope receiver. If there is some antenna pattern mismatch with the radio telescope, the beacon power must be adjusted accordingly to compensate for the difference in uplink versus downlink attenuation. 

\paragraph{Placement of beacon transmitters} Beacon transmitters can be located either at or nearby the radio telescope. Co-locating the beacon transmitter at the radio telescope may pose dynamic range limitations to avoid causing compression or damage to the radio telescope front-end. 
Instead, the beacon transmitters may be placed around the radio telescope. In that scenario, the radio telescope and the transmitters can communicate with each other through a separate dedicated (perhaps wired) channel. The drawback is potential mismatch between antenna patterns associated with the beacons and radio telescope.

\subsection{Satellite Transmissions}\label{subsec:trans_schemes}

When a satellite enters the field-of-view of the radio telescope, we assume that the satellite collects the beacon energy by integrating the received signal over the designated time-frequency-spatial pattern assigned to the beacon. Referring to Fig. \ref{fig:beacon_sensing}, we denote the duration of this energy collection as $T_b$, a design parameter. This must be less than the travel time over the field-of-view of the radio telescope. To exploit reciprocity, $T_b$ should also be less than the coherence time of the channel between the satellite and the radio telescope. 
%If $T_b$ is sufficiently large, 
The satellites must not transmit while listening the beacon signal. 
We present two alternative scenarios for controlling satellite transmissions.\footnote{These schemes could be combined with additional signaling about the measurement schedule over the command and control channel used to manage the satellite constellation.}
%\dg{SO IF Tb IS SMALL, SATELLITES DO NOT TURN OFF? WHY?}

 First, the satellite senses the beacon only once when the satellite first enters the field-of-view of the radio telescope. If the beacon energy collected at the satellite exceeds a certain threshold, the beacon signal is detected at the satellite. The threshold depends on the satellite receiver sensitivity. Once the beacon signal is detected, the satellite ceases to transmit until it exits the field-of-view of the radio telescope. If the beacon signal is not detected at the satellite,  the satellite transmits until it exits the field-of-view of the radio telescope. However, the RFI generated by the satellite may increase during its travel over the field-of-view of the radio telescope, increasing the probability of excessive RFI.

Second, the satellite periodically senses for the presence of the beacon signal. The satellite disrupts transmissions to periodically sense the beacon with period $T_p$. The first $T_b$ seconds of each period are reserved for measuring the energy of the beacon so that $T_b/T_p$ represents the overhead spent on sensing. If the beacon is detected in any of those intervals, then the satellite ceases to transmit, but continues to sense the beacon, and resumes transmissions if the beacon is no longer detected.  If the satellite does not detect the beacon, its transmitter turns on and off according to the sensing duty cycle.

%\mh{I changed this since it seems to me this matches your analysis -- we may need to discuss.}
%\dg{IS THIS HOW YOU SIMULATE? ALSO, IF THE SATELLITE LISTENS PERIODICALLY, CAN IT TURN BACK ON ONCE IT STOPS DETECTING THE BEACON?}

\section{Performance Evaluation}
\label{sec:NumRes}

\subsection{Assumptions} 
We assume a single radio telescope with a fixed look direction over the measurement interval. The arrival and departure processes of the satellites across the field-of-view of the radio telescope are stationary processes. We assume that the satellite power profiles are %uniform over frequency and
constant over time when active. %For the quiet zone methods, 
The process $\{\RFI(t)\mid t \in[0, T]\}$ is assumed to be stationary. 

For the beacon approach, we assume that there is a single beacon transmitter located at the radio telescope. In addition, we assume that the satellites periodically sense the beacon signal, and that the measurement interval for the radio telescope is much larger than the beacon period $T_p$, as described in Sec.~\ref{subsec:trans_schemes}. At any given time $t$ outside of the sensing period $T_b$, any satellite located within the field-of-view is transmitting with some probability based on whether or not it detected the beacon in the preceding sensing interval. 

Since $\{\RFI(t)\mid t \in[0, T]\}$ is assumed stationary, 
%\dg{STATIONARITY DOES NOT IMPLY ergodicity, SO THE LOGIC IS NOT SOUND HERE. ALSO, WHY SAY ergodic A SECOND TIME AFTER ONCE IN THE LAST PARAGRAPH? ACTUALLY, I DON'T SEE WHY ergodicity IS NEEDED. IS STATIONARITY NOT ENOUGH?}
we consider a snapshot of the network.  That is, we drop the dependence on $t$. %The outage probability in \eqref{eq:outage} reverts to
% \begin{align}
%     P_{\text{out}} = \Pr\{\RFI\ge \RFImax\}.
% \end{align}
We assume that the network consists of $N$ satellites where their locations are uniformly distributed over the surface of a sphere centered at $(0, 0, 0)$ with radius $R_e + h~$km.\footnote{This is an approximation of the actual distribution since satellites are launched in particular orbits and inclinations.} Specifically, the locations are generated according to the homogeneous binomial point process \cite[Prop.~1]{wang2022evaluating}.  If $M$ of these satellites lie in the field-of-view of the radio telescope, we assume $M$ ground stations uniformly distributed over a region on the earth. In this region, each point is located at a distance of less than $100~$km from the radio telescope. $M$ is a random variable dependent on the realization of the satellite locations. The associations between the satellites and ground stations are selected to minimize the total distance between them. The transmit power of each active satellite is $p_s$.  
The short term channel variations are modeled as log-normal random variables, $10\log_{10}(\xi_i)\sim\mathcal{N}\left(0, \sigma_{\text{dB}}^2\right)$ for each $i$.

Deactivation of the satellite links in the quiet zone approach depend only on their locations. For the beacon approach, on average $MT_b/T_p$ satellites are listening for the beacon signal, and the remaining $M(T_p-T_b)/T_p$ satellites have made their decisions based on the preceding listening interval so that they are either transmitting or are deactivated. For the simulations we assume that $T_b/T_p$ is small so that approximately all satellites have been instructed to transmit or remain silent. %\mh{I don't understand this since all active satellites listen to the beacon during $T_b$, and the rest are shut off. If we consider t outside the beacon interval then I thought your analysis can be applied to all satellites assuming they turn on/off according to whether or not they detected the beacon in the last period.}

%\co{Cuneyd: If a satellite listens to the beacon, then it is off. For any given snapshot, for a given satellite, there are two possibilities:  the satellite is listening to the beacon, or the satellite already made its decision based on the beacon energy. If there are N many satellites above the horizon, then on the average $NT_b/T_p$ of them  are listening the beacon signal, so they are off. Others, $N(T_p-T_b)/T_p$ of them, made their decisions. We need to consider the ones made their decision. If $T_b/T_p$ is small, we can approximate the all satellites are already listened the beacon, and made their decision. We are evaluating the RFI based on their decisions.}
%\mh{my understanding is that during $T_b$ *all* satellites in the field-of-view are turned off. those that decide to transmit must listen again for the next beacon. no?}
%\co{Those that decide to transmit will listen again. But those time slots changes for each satellite. I mean a satellite listens the beacon when it enters the horizon. In other words, satellites can listen the beacon signal in different time slots. They are not synchronized. However, as I explained above, at any given time, some of them are listening, some of them already made their decision.}

%\mh{I don't see how they are not synchronized with the beacon transmissions -- let's discuss tomorrow}
%\co{Sure Professor. The deadline is extended by the way.}

% yes I saw that, we have more time

\subsection{Dynamic Quiet Zones versus Beacons}
Let $\gamma_i$ be the channel gain between the $i$-th satellite and the radio telescope. For the dynamic quiet zone approach, the aggregated RFI at the radio telescope can be written as
\begin{align}
    \RFI_{\text{d}} = \sum_{i=1}^{N} p_s \gamma_i \xi_i \mathbbm{1}\{p_s\gamma_i \leq \tau\},\label{eq:varying_quiet_rfi}
\end{align}
where $\tau$ is the $\RFI$ threshold level. 

 $T_b$ is assumed to be shorter than the coherence times of the satellite channels. The beacon is transmitted over the frequency band $B_b$ with power $p_b$ for $t\in[0, T]$. Let $\widetilde{\gamma}_i$ denote the channel gain for the beacon transmission to the $i$-th satellite. If there is no antenna mismatch, $\widetilde{\gamma}_i = \gamma_i$. Let $\eta$ denote the threshold ratio of received beacon energy to noise density $N_0$ at the satellite. That is, the beacon is detected at the $i$th satellite if $p_bT_b\widetilde{\gamma}_i\xi_i\ge \eta N_0$.

%The beacon signal is transmitted at the radio telescope using the receiving antenna pattern with a different maximum gain

% \dg{SUGGEST WRITING THE FOLLOWING IN PLAIN ENGLISH RATHER THAN USING $\implies$.}
% \begin{align}
%     \frac{p_bT_b\gamma_i\xi_i}{N_0} \geq \eta \implies \text{the beacon is detected at the $i$-th satellite.}
% \end{align}
Therefore, the aggregate RFI at the radio telescope with the beacon-assisted protocol is
\begin{align}
    \RFI_{\text{b}} = \sum_{i=1}^{N} p_s \gamma_i \xi_i \mathbbm{1}\{p_bT_b\widetilde{\gamma}_i \xi_i < \eta N_0\} \label{eq:beacon_rfi}.
\end{align}
%\dg{IT'S CONFUSING TO USE ``RFI'' IN (6) AND (8) TO REPRESENT TWO DIFFERENT VALUES. IN GENERAL, USE DIFFERENT NOTATIONS FOR DIFFERENT THINGS.}
From \eqref{eq:varying_quiet_rfi} and \eqref{eq:beacon_rfi}, if $p_b$ and $T_b$ are chosen such that
 \begin{align}
     \frac{\eta N_0}{p_b T_b} = \frac{\tau}{p_s}, \label{eq:vary_quiet_vs_beacon}
 \end{align}
  RFI levels are deterministic, i.e., $\xi_i = 1$ for all $i$, and there is no antenna mismatch between the beacon and the radio telescope, then the dynamic quiet zone and beacon-assisted approaches turn off the same satellite links and achieve the same aggregate $\RFI$ level ($\RFI_{\text{d}} = \RFI_{\text{b}})$.

 \subsection{Simulations}
  \begin{table}
    \centering
    \begin{tabular}{|c|c|}
    \hline
       Carrier frequency  & $10.65$~[GHz] \\
       \hline
       Bandwidth & $100$~[MHz]\\
       \hline
       $\text{PFD}_{\text{max}}$ & $-240$~[dBW/m$^2$/Hz]\\
       \hline
       Satellite altitude & $550$~[km] \\
       \hline
       Satellite transmit power  & $-8.3$~[dBW/MHz]\\
       \hline
       Maximum gain of the satellite antenna  & $30$~[dBi]\\
       \hline
       Satellite beam-width & $5$~[degree]\\
        \hline
       Maximum gain of the radio telescope ($G_{\text{max}}$) & 64~[dBi]\\
        \hline
        Beacon power $(p_b)$  & 10~[mW] \\
        \hline
        Maximum gain of the beacon transmitter & 32~[dBi]\\
        \hline
        Threshold at the satellite $(\eta)$ & 9.6~[dB]\\
        \hline
        $\sigma_{\text{dB}}$ & 5~[dB]\\
        \hline
        Sky (Antenna) noise temperature ($T_a$) & 300~[K]\\
        \hline
        Satellite receiver noise temperature ($T_r$) & 100~[K]\\
        \hline
         Radio telescope antenna pattern & ITU-RA 1631 \cite{ITURA1631} \\
        \hline
        Beacon antenna pattern & ITU-RA 1631 \cite{ITURA1631} \\
        \hline
        Satellite antenna pattern & 3GPP TR 38.811  \cite{3GPP38811}\\
        \hline
    \end{tabular}
\vspace{0.5cm}
\caption{Satellite and radio telescope parameters.}
\label{table:1}
\end{table}
\begin{figure}
    \centering
    \includegraphics[width = \linewidth]{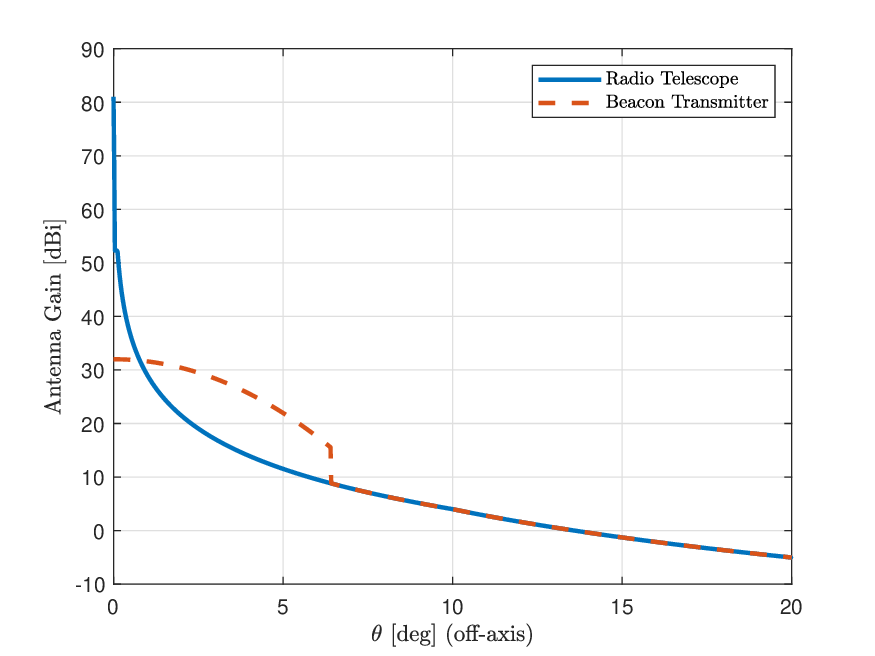}
    \caption{Antenna gain versus off-axis angle for the radio telescope and the beacon transmitter.
    }
    \label{fig:antenna_mismatch}
\end{figure}
\begin{figure}
    \centering
    \includegraphics[width = \linewidth]{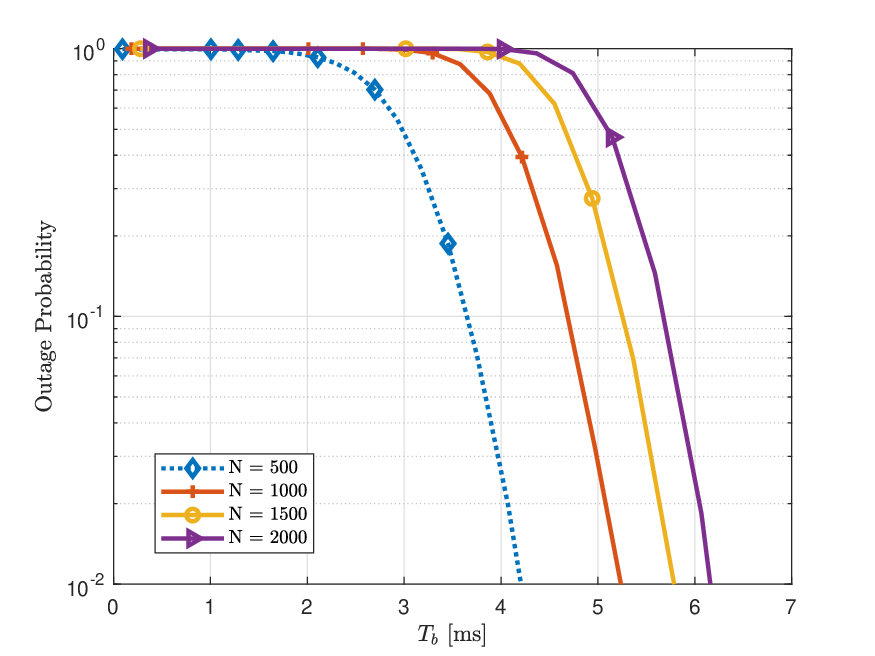}
    \caption{Outage probability versus $T_b$ for the beacon approach for $N\in\{500, 1000, 1500, 2000\}$.}
    \label{fig:outage_vs_Tb}
\end{figure}
\begin{figure}
    \centering
    \includegraphics[width = \linewidth]{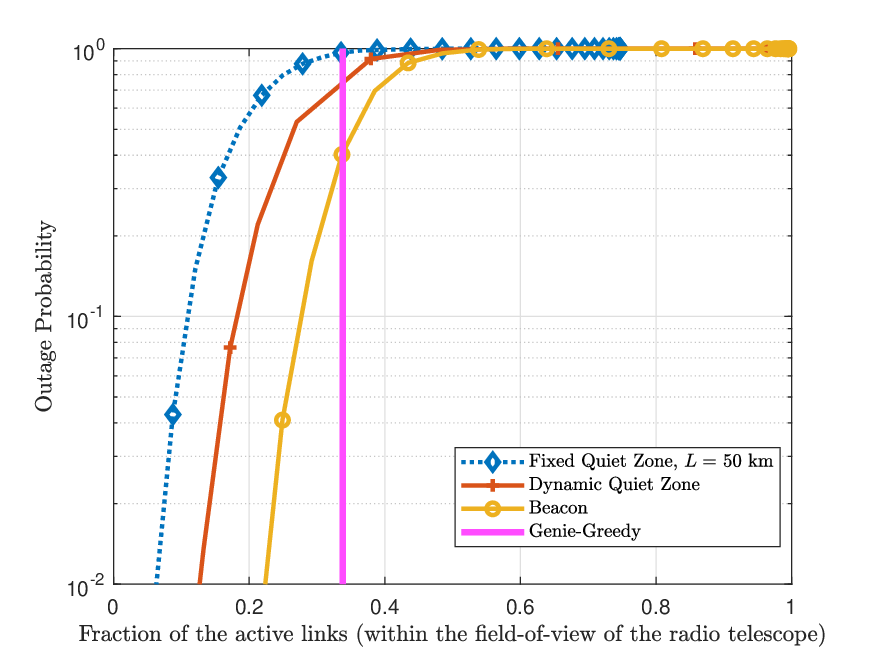}
    \caption{Outage probability versus the fraction of the active links for the fixed quiet zones, the dynamic quiet zones, the beacon approach and the genie-aided approach, when $N = 1000$.}
    \label{fig:outagevsactivelink_N=1000}
\end{figure}
The simulation parameters are given in Table~\ref{table:1}. We consider the frequency band $10.6-10.7$~GHz centered around $f_c = 10.65$~GHz, which is assigned to radio astronomy. %For that band, the typical maximum antenna gain $(G_{\text{max}})$ of the radio telescope is $6~$dBi \cite{ITURA1631}. 
The radio telescope antenna diameter $D$ can be found from
 \begin{align}
    G_{\text{max}} = 20\log_{10}  \left(\frac{\pi D}{\lambda}\right),
 \end{align}
 where $\lambda$ is the wavelength\cite{ITURA1631}. The threshold for the power flux density
 $(\text{PFD}_{\text{max}})$ is taken from \cite{2003_ITU_RA_769} for $T = 2000$~seconds. The relationship between $\text{PFD}_{\text{max}}$ and $\RFImax$ is given by
 \begin{align}
     \RFImax =  \text{PFD}_{\text{max}} \pi (D/2)^2\times \text{Bandwidth}.
 \end{align}
The noise power spectral density at the satellite, $N_0$, is given by $N_0 = k (T_a + T_r)$, where $k$ is the Boltzmann constant $(1.38\times 10^{-23} \text{J/K})$, and $T_a$ and $T_r$ are sky noise temperature and satellite receiver noise temperature, respectively.

For the beacon transmission, we assume an antenna mismatch with the radio telescope corresponding to a reduction in antenna gain: the maximum gain of the beacon transmitter is assumed to be $32$~dBi whereas for the radio-telescope it is 64 dBi. The mismatch in the resulting patterns is shown in Fig.~\ref{fig:antenna_mismatch}. The figure shows that the mismatch goes to zero when the off-axis angle is greater than $6.5^\circ$.

 % Then, the genie simply find the \co{largest} $K$ (if there is) such that \dg{CONFUSING. AT MOST ONE K CAN SATISFY THE FOLLOWING? WHY smallest?}

%\dg{SUGGEST USING MARKERS IN FIGURE TO DISTINGUISH CURVES IN BLACK-WHITE PRINT.}
Fig.~\ref{fig:outage_vs_Tb} shows outage probability for the beacon approach versus $T_b$ for $N\in\{500, 1000, 1500, 2000\}$. The corresponding average numbers of satellites in the field-of-view are $\{21, 42, 63, 84\}.$ It is evident that an increase in the number of satellites necessitates a longer beacon duration $T_b$ in order to maintain a consistent outage probability. It is important to note that the value of $T_b$ is contingent upon the power of the beacon. Specifically, a decrease in beacon power requires an increase in the duration of $T_b$.

Fig.~\ref{fig:outagevsactivelink_N=1000} compares the performance of the quiet zones (fixed and dynamic) with the beacon approach for $N=1000$. For this example, the average number of satellites in the field-of-view is $42$. The figure shows outage probability versus the fraction of active satellite links each method. The fraction of active satellite links are computed for the satellites located within the radio telescope's field of view. 

Also shown as a performance benchmark is the performance of a genie-aided greedy algorithm, where the genie knows the RFI generated by each satellite and deactivates the links associated with the largest RFI to ensure that $\RFI < \RFImax$. 
Specifically, for the genie-aided method let $(I_i)_{i=1}^{N}$ denote the interference levels of the satellite links where $I_i\le I_j$ when $i\le j$. If $\sum_{i=1}^{N} I_i \leq \RFImax$, all links remain active, and if $I_1\geq \RFImax$, all links are deactivated. Otherwise, there exists a unique $K\in\mathbb{N}$ such that $1\le K < N$ and 
 \begin{align}
     \sum_{i=1}^{K} I_i \le \RFImax \le \sum_{i=1}^{K+1} I_i.
 \end{align}
  Then, the first $K$ links are active and the others are deactivated. Hence, in the genie aided approach, the outage probability is zero.  

%\mh{I added a bit more detail -- check for accuracy.} 
To generate the simulation results in Fig.~\ref{fig:outagevsactivelink_N=1000}, in each random trial the satellites and the ground stations are dropped uniformly over the spherical region. 
%\mh{correct? this may not be realistic due to curvature...}  %\co{Cuneyd: Yes, correct. We ignore the curvature.} 
For the fixed quiet zone $L = 50~$km, both the outage probability and the fraction of the active links within the radio telescope's field of view are computed for a range of quiet zone radii in the sky $V$ between 0 and $2\sqrt{h(2R_e+h)}$. That is, the fixed quiet zone curve in Fig.~\ref{fig:outagevsactivelink_N=1000} is obtained for $L = 50$~km and all quantized values of $V$ in that range. 
For the dynamic quiet zone curve, the corresponding outage probabilities and the fractions of the active links are computed for different values of the RFI threshold $\tau\in[10^{-17}, 10^{-14}]$. Similarly, the results for the beacon are obtained for values of $T_b\in[0, 5.3]$~ms.

% I left a note in the Fig 8 caption -- minor suggestion. 
% I'm signing off for now, may check back in a bit

%each point in  corresponds to and  value for $V$ in that range.  
% I don't understand -- I'll leave it to you to fix it as you think is best.

%not the best value, all possible V values in this range. the best value yields a single point, the curve is obtained over all $V$ values in this range.
%Not an optimized value of $V$. Each point in this curve corresponds to a (L, V) pair, L = 30 is fixed, the curve is obtained for different values of $V$. $V$ can take values between 0 and the maximum distance in the field-of-view, $

%For the fixed quiet zone $L= 30~$km, and the outage probability is minimized over possible quiet zone radii in the sky $V$. 
%During the process of obtaining the curves showing the outage probability with respect to the fraction of active links, we employ the following methodology. The computation of both the outage probability and fraction of active links is performed for the fixed quiet zone curve with $L = 30~$km, encompassing all potential values of the quiet zone radius in the sky ($V$). 

%Similarly, the dynamic quiet zone results correspond to the minimum outage probability over feasible values of the RFI threshold ($\tau$). For the beacon approach the results correspond to the minimum over $T_b$. 

Even with the genie-aided approach,  a minimum of $66\%$ of the satellites located within the field-of-view of the radio telescope must be deactivated to ensure that $\RFI < \RFImax$. 
%In other words, for the given parameters the absence of precautionary measures are likely to lead to a high probability of outage. %
The results show that for an outage probability $\le 10^{-2}$, dynamic quiet zones enable a significant increase in active satellite links relative to fixed quiet zones, and the beacon approach provides a similar further gain relative to the dynamic quiet zones. The gain depends on the variance of $\xi_i$, which represents the random channel variations as $\xi_i$ increases, so does the relative gain of the beacon-assisted approach. 
%It is anticipated that an increase in short-term channel variations will result in a further increase in the gain attributed to the beacon approach. 

In addition, we expect that the gap between the beacon and genie-aided approaches decreases as $N$ decreases. In particular, when the radio telescope's field-of-view encompasses only a single satellite and there is no mismatch between the antenna patterns, the beacon-assisted performance achieves the genie-aided bound.

%\mh{N decreases? $T_b$ increases? can the gap go to zero? if not, why not?}
%

%\co{Cuneyd: We do not have any graph showing outage vs active link fraction for different $N$ and $T_b$s, I can add those graphs, if you think it is important.}
% I am asking if you can predict this -- if it's difficult, then maybe we need more plots, but we should be able to say in general under what circumstances we expect the gaps to close

%Moreover, it can be inferred that by employing the dynamic quiet zone approach, it is possible to enable increase the number of links compared to the fixed quiet zone approach in order to achieve an outage level of $10^{-3}$. Furthermore, by utilizing the beacon approach, it is possible to enhance performance by leveraging the instantaneous channel information. 

%\mh{when can we expect the gap between dynamic QZ and beacon to close? for what parameter range?}
%\co{Clearly, when $N$ decreases, the gap decreases, too. When $T_b$ increases, I am not sure.}

\section{Conclusions}
We have proposed a beacon-assisted protocol to facilitate sharing of spectrum between passive sensing and a satellite network. The beacon exploits channel near reciprocity, i.e., if the beacon energy received at a satellite exceeds a threshold then the satellite is likely to generate excessive RFI, and is then instructed not to transmit. The beacon power can be adjusted to ensure that the aggregate RFI from multiple satellites does not exceed a threshold that disrupts measurements at the passive sensor. For parameters selected based on a network of LEO satellites, numerical results show that a beacon with modest power combined with integration times of a few msecs can reduce the probability of excess RFI (outage probability) to less than 0.01. When compared with a conventional fixed quiet zone surrounding the radio-telescope, the beacon offers a significant increase in the number of active satellite links that can be supported with the same outage probability. In the presence of random channel variations (e.g. due to changing weather conditions), the beacon approach can also offer a significant gain with respect to a dynamic scheme for defining quiet zones based on predictable satellite trajectories.  %, then the throughput of the satellite network We have employed low-energy beacon signal with a smaller maximum transmitter antenna gain than the maximum antenna gain of the radio telescope. 
%It has been illustrated that a greater number of satellite links compared to the RQZ approaches. 

Although we have focused on the scenario in which a radio-telescope shares spectrum with a satellite network, the general approach applies to spectrum sharing between other types of passive sensors and terrestrial wireless networks. An extension of this work is to consider the design of the beacon signal with shorter integration times at the passive sensor. In that scenario the outage probability should take into account nonstationarities in received RFI. Another extension is to consider the design of a distributed set of beacon transmitters nearby the radio telescope or transmitters directed radially around a passive sensor. Such approaches could  communicate with terrestrial networks or user devices. 
%Furthermore, the beacon-assisted methodology has the potential to facilitate coexistence between radio astronomy and other active networks, including terrestrial communication networks.

%We have discussed system considerations associated with the beacon design and potential schemes for the beacon transmission.A numerical comparison has been conducted to evaluate the performance of the beacon approach in relation to the two versions of the RQZs. 

% It has been shown that even though there is an antenna mismatch between the radio telescope and the beacon transmitter, via the beacon approach more satellite links may be enabled comparing to the RQZ approaches. 

\bibliographystyle{IEEEtran}
\bibliography{references}

\begin{thebibliography}{10}
\providecommand{\url}[1]{#1}
\csname url@samestyle\endcsname
\providecommand{\newblock}{\relax}
\providecommand{\bibinfo}[2]{#2}
\providecommand{\BIBentrySTDinterwordspacing}{\spaceskip=0pt\relax}
\providecommand{\BIBentryALTinterwordstretchfactor}{4}
\providecommand{\BIBentryALTinterwordspacing}{\spaceskip=\fontdimen2\font plus
\BIBentryALTinterwordstretchfactor\fontdimen3\font minus
  \fontdimen4\font\relax}
\providecommand{\BIBforeignlanguage}[2]{{%
\expandafter\ifx\csname l@#1\endcsname\relax
\typeout{** WARNING: IEEEtran.bst: No hyphenation pattern has been}%
\typeout{** loaded for the language `#1'. Using the pattern for}%
\typeout{** the default language instead.}%
\else
\language=\csname l@#1\endcsname
\fi
#2}}
\providecommand{\BIBdecl}{\relax}
\BIBdecl

\bibitem{2021_ITU_R_2259}
ITU-R, ``Characteristics of radio quiet zones,'' \emph{RA.2259-1}, 2021.

\bibitem{2014_Carol_PropPrediction}
C.~Wilson, ``Propagation prediction in establishing a radio quiet zone for
  radioastronomy,'' in \emph{The 8th European Conference on Antennas and
  Propagation (EuCAP 2014)}, 2014, pp. 1209--1213.

\bibitem{2021_Minn_NGSO_Ras}
Z.~Fan, Y.~Dai, and H.~Minn, ``Performance analysis of large-scale ngso
  satellite-based radio astronomy systems,'' \emph{IEEE Access}, vol.~9, pp.
  93\,954--93\,966, 2021.

\bibitem{2023_Chakraborty_CellularRFI}
S.~Chakraborty, G.~Hellbourg, M.~Careem, D.~Saha, and A.~Dutta, ``Collaboration
  with cellular networks for rfi cancellation at radio telescope,'' \emph{IEEE
  Transactions on Cognitive Communications and Networking}, pp. 1--1, 2023.

\bibitem{2022_Weldegebriel_Pseudonymetry}
M.~G. Weldegebriel, J.~Wang, N.~Zhang, and N.~Patwari, ``Pseudonymetry:
  Precise, private closed loop control for spectrum reuse with passive
  receivers,'' in \emph{2022 IEEE International Conference on RFID (RFID)},
  2022, pp. 91--96.

\bibitem{2013_Bryerton_Cyrogenic}
E.~W. Bryerton, M.~Morgan, and M.~W. Pospieszalski, ``Ultra low noise cryogenic
  amplifiers for radio astronomy,'' in \emph{2013 IEEE Radio and Wireless
  Symposium}, 2013, pp. 358--360.

\bibitem{2022_ITU_R_2188}
ITU-R, ``Power flux-density and e.i.r.p. levels potentially damaging to radio
  astronomy receivers,'' \emph{RA.2188-1}, 2022.

\bibitem{2023_Zheleva_RDZ}
M.~Zheleva, C.~R. Anderson, M.~Aksoy, J.~T. Johnson, H.~Affinnih, and C.~G.
  DePree, ``Radio dynamic zones: Motivations, challenges, and opportunities to
  catalyze spectrum coexistence,'' \emph{IEEE Communications Magazine}, pp.
  1--7, 2023.

\bibitem{2023_Minn_SRQZ}
H.~Minn and Z.~Fan, ``Sky radio quiet zones for mitigating rfi from large-scale
  ngso satellites to ground radio astronomy system,'' \emph{IEEE Access},
  vol.~11, pp. 91\,336--91\,357, 2023.

\bibitem{2014_Altamimi_Enforcement}
M.~Altamimi and M.~B. Weiss, ``Enforcement and network capacity in spectrum
  sharing: Quantifying the benefits of different enforcement scenarios,''
  \emph{Available at SSRN 2481082}, 2014.

\bibitem{2016_Minn_SpecSharingCellular}
H.~Minn, Y.~R. Ramadan, and Y.~Dai, ``A new shared spectrum access paradigm
  between cellular wireless communications and radio astronomy,'' in \emph{2016
  IEEE Global Communications Conference (GLOBECOM)}, 2016, pp. 1--6.

\bibitem{cakaj2021parameters}
S.~Cakaj, ``The parameters comparison of the “starlink” leo satellites
  constellation for different orbital shells,'' \emph{Frontiers in
  Communications and Networks}, vol.~2, p. 643095, 2021.

\bibitem{2019_ITU-P_525}
ITU-R, ``Calculation of free-space attenuation,'' \emph{P.525-4}, 2019.

\bibitem{wang2022evaluating}
R.~Wang, M.~A. Kishk, and M.-S. Alouini, ``Evaluating the accuracy of
  stochastic geometry based models for leo satellite networks analysis,''
  \emph{IEEE Communications Letters}, vol.~26, no.~10, pp. 2440--2444, 2022.

\bibitem{ITURA1631}
ITU-R, ``Reference radio astronomy antenna pattern to be used for compatibility
  analyses between non-gso systems and radio astronomy service stations based
  on the epfd concept.'' \emph{Recommendation ITU-RA.1631}, 2003.

\bibitem{3GPP38811}
3rd Generation Partnership Project; Technical Specification Group~Services and
  S.~Aspects, ``Study on new radio (nr) to support non terrestrialnetworks;
  (release 15),'' \emph{3GPP TR 38.811 V1.0.0}, 2018.

\bibitem{2003_ITU_RA_769}
ITU-R, ``Protection criteria used for radio astronomical measurements,''
  \emph{RA.769-2}, 2023.

\end{thebibliography}
	
\end{document}